\documentclass[12pt]{article}

\usepackage{amssymb}
\usepackage{amsfonts}
\usepackage{amsmath}
\usepackage{amsthm}

\title{Hertz's viewpoint on quantum theory}

\author{Andrei Khrennikov\\International Center for Mathematical Modeling \\
in Physics, Engineering, Economics, and Cognitive Science\\
Linnaeus University, V\"axj\"o, Sweden  }

\begin{document}
\maketitle

\begin{abstract}
In 19th century (in the process of transition from mechanics to the field theory)  the German school of theoretical physics confronted problems similar 
to the basic problems in the foundations of quantum mechanics (QM). Hertz tried to resolve such problem through analysis of the notion of a scientific theory and interrelation 
of theory and experiment.   This analysis led him to the Bild (image) conception of theory (which was latter essentially developed, but also modified by Boltzmann). 
In this paper we claim that   to resolve the basic foundational problems of QM, one has to use the Bild conception and reject the observational viewpoint on physical theory.
As an example of a Bild theory underlying QM (treated as an observational theory), we consider   prequantum classical statistical field theory (PCSFT): theory of random subquantum 
fields. 
 \end{abstract}

{\bf keywords:} Hertz Bild theory, descriptive and observational theories, hidden variables in electromagnetism vs quantum mechanics, quantum theory as observational theory,
prequantum classical statistical field theory.  
 
 \section{Introduction}
 
 During one hundred years quantum theory has been suffering of endless debates about its meaning and interpretation. 
 I claim that this unacceptable situation is the result of neglect  by the fathers of quantum mechanics (QM) the  extensive study of similar problems by the traditional German school in physics: 
ignoring the works of Helmholtz, Hertz, and Boltzmann, see, e.g.,  (Hertz, 1899; Boltzmann, 1905, 1974).  Consciously or unconsciously  Bohr, Heisenberg, Einstein, Pauli and other main contributors
to foundations of  quantum theory (but excluding Schr\"odinger, see, for example, (D' Agostino, 1992)) 
ignored the historical lessons of the debate on the interrelation between theory and experiment which was initiated by transition 
from Newtonian mechanics to Maxwellian electromagnetism  (Hertz, 1899).  In particular, in this debate the problem of hidden variables was enlighten by Hertz - may the first time 
in history of science   (Hertz, 1899).  In the light of this debate the following debate between Bohr and Einstein can 
be characterized by lack of deep philosophic analysis (Einstein, Podolsky, \& Rosen, 1935; 
Bohr, 1935).  I am not afraid to call  the latter debaters naive - by taking into account the lessons of the aforementioned debate about electromagnetism.

In this paper I shortly present the views of Hertz, see, e.g.,  (Hertz, 1899), see also (Boltzmann, 1905, 1974), on scientific theory  - the {\it  Bild  (image) conception}, section \ref{BC}.   
Here I follow the  works (D' Agostino, 1992;  Miller, 1984). At the end of this section there are discussed various 
approaches to the notion of theory, {\it the descriptive,  Bild, and observational approaches.} By speaking about a scientific theory 
one has to specify its type.  Then I proceed to quantum physics. I consider the present situation in quantum foundations by appealing to the 
Herzian Bild conception, section \ref{HQM}.
In section \ref{CC} there are formulated the rules of correspondence between two theories of different types (especially their
 mathematical structures). 
\footnote{This paper is continuation of my previous works in which the Bild-conception was explored in quantum physics (Khrennikov, 2017a, b, c).  
It is also important to remark that similar approach to quantum theory was supported by Schr\"odinger, see (D'Agostino, 1992).} 
Finally, in section \ref{PQ} there is presented a theory of micro-phenomena based on the Bild conception, {\it prequantum classical statistical field theory} (PCSFT), 
see (Khrennikov,   2007a, b, 2014, 2017c).   

Since this issue is devoted to ontology of quantum theory, it is useful to stress the impact of the Bild conception to the quantum foundational 
 debate about realism, including Einstein-Bohr debate about completeness of QM.  From the Bild-viewpoint, realism in physics as well as any 
other area of scientific research is reduced to experimental facts. This is exactly Bohr's position (see, e.g.,  Plotnitsky, 2006, 2009).\footnote{However, Bohr would say that 
a Bild-type theory has nothing to do with physics and he would  refer to it as a metaphysical theory. At the same time he was not so much interested in no-go theorems for descriptive 
or Bild-type theories. In principle, he could not exclude that such `beyond quantum theories' might be constructed. But, they would not have any value for physics  (Plotnitsky, 2006, 2009). } Thus the only realistic component 
of quantum physics are outcomes of measurements (Bohr's `phenomena'). Any physical theory is only about human images of natural phenomena. At the same time these images 
are created on the basis of human's interaction with nature. 

In (Khrennikov, 2007c) I tried to establish relation between the Hertz-Boltzmann Bild viewpoint and the ontic-epistemic viewpoint (Atmanspacher  \& Primas, 2005)  
on the notion of scientific theory. However, this is a complex problem. Observational theories of the present paper can be definitely treated as epistemic theories. 
However, the Bild conception is not about reality as it is (as in an ontic theory), it is about human images of reality.

\section{Method}

\subsection{Bild conception}
\label{BC} 

Hertz' discovery of radio-waves was connected with his deep analysis of the Maxwellian electromagnetism from the viewpoint of interrelation between theory and experiment.  
Electromagentism, in Hertz's opinion, based on {\it the action at a distance principle} was only a  ``first approximation to the truth.'' And he worked hardly to approach the 
final true theory.  From the formal viewpoint, he tried to create a mechanical model of electromagnetic phenomena. However, these studies led him to understanding that
it seems to be impossible to construct such a model without invention of {\it hidden variables} of the mass type, so called  concealed masses. In turn, this led him to  deep philosophical 
and methodological studies devoted to meaning of `theory' in science.\footnote{He was not able to complete his project on the mechanical theory of electromagnetism. 
(Ironically the same fate befell Einstein who in turn spent the last 20 years of his life by attempting to create the classical field theory of quantum phenomena, see, e.g., 
(Einstein \& Infeld, 1961).) However, 
Hertz's contribution was very valuable to the methodology of science. And it influenced strongly Boltzmann and Schr\"odinger and through Botzmann's works Planck (and may be 
even Einstein), see (Miller, 1984).}    

The main impact of these studies was relative liberation of theory from experiment. One of Hertz' fundamental statements is that 
{\footnotesize\em ``We become convinced that the manifold of the actual universe must be greater than the manifold of the universe which id directly revealed to our
senses.''}  See (Hertz, 1899). 

Hertz explored heavily Helmholtz's  principle about a parallelism between concepts and perceptions. 
However, Hertz rejected  Helmholtz's  claim that this parallelism uniquely determines the theory consistent 
with experimental facts. Hertz questioned the later (so to say the strong version 
of the  parallelism principle)  and claimed that there exists a multiplicity of representations satisfying the requirement of Helmholtz's parallelism: 
{\footnotesize\em ``The images [Bilder] which we may form of things are not determined without ambiguity by the 
requirement that the consequents of images must be images of consequents. Various images of the 
same objects are possible, and these images may differ in various aspects.''} See (Hertz, 1899).

It is even more important for our present considerations that Hertz stated that Helmholtz's parallelism of laws does not even work if 
{\it a theory is limited to visible quantities.} Only the introduction of {\it hidden quantities} allows
 creation of a consistent theory:
{\footnotesize\em ``If we try to understand the motions of bodies around us, and refer to simple and clear rule, paying attention only 
to what can be {\it directly observed}, our attempts will in general fail. We soon become aware that 
the totality of things visible and tangible do not form a   universe conformable to law, in which the same 
result always follow from the same conditions.'' } See (Hertz, 1899).

From Hertzian perspective, a theory is not a true description of nature (Botzmann's ``complete congruence with nature'') or at least a best approximation of it, but 
a theory is ``mere a representation (Bild)   of a nature ... which at the present  allows 
one to give the most uniform and comprehensive account of totality 
of phenomena'' See (Boltzmann, 1905).

Of course, this viewpoint on the conception of theory represents  a failure from the perspective of the traditional descriptive 
conception of theories. However, this liberation of theory from experiment has its big advantage, since it liberates  scientist's mind 
from rigid constraints of the present experimental  situation. 

Treatment of theory as a consistent system of mental images leads to its {\it causality.} The latter is a consequence of causality of human reasoning. In his reasoning 
a human cannot do anything else than to proceed from cause to its consequence. At the same time causality should not be treated as a purely mental (logical) feature of a theory.
We recall that Helmholtz's parallelism between sensation and perception played the fundamental role in establishing of the Bild conception. Therefore the causal structure 
of human reasoning is the result of evolutionary experiencing of humans observing causality in natural processes.

Since the Hertz(-Botzmann) viewpoint on the notion of theory has not been commonly accepted, it is useful to specify it by calling 
{\it `Bild-theory'.} It should be distinguished from  {\it `descriptive theory'}  attempting to provide ``complete congruence with nature.'' 
Besides Bild and descriptive theories, we consider {\it `observational theory'} operating only with outputs of observations. This sort of theory 
can also be called {\it `sensational theory'}, in contrast to Bild theory which can be called {\it `perceptional theory'.} The same experimental situation 
can be represented by various types of theories: descriptive, Bild, and observational. 

\subsection{Hertzian viewpoint on foundations of quantum mechanics}
\label{HQM}

For our considerations, the most important is that  Hertzian analysis of methodology of science implies:
\begin{enumerate}
\item Any attempt to create a consistent (causal)  theory on the purely experimental basis would lead to  a failure;    
\item Any consistent theory of natural phenomena would contain hidden variables, quantities which are unapproachable 
for our perception (at least at the present time); 
\item Generally in theory it is impossible to approach the one-to-one correspondence between
 theoretical concepts and experimental facts.
\end{enumerate}   

We state that these principles were totally ignored not only by fathers of QM (with a few exceptions such as Schr\"odinger), but even by practically 
all experts working in quantum foundations. The majority of them followed `the spirit of Copenhagen' (Plotnitsky, 2016)  and put tremendous efforts to proceed without taking 
into account Hertz 1, 2, i.e., to develop the formalism of observational (sensational)  theory of micro-phenomena which is nowadays known as QM (cf. with Stapp's analysis of 
the Copenhagen interpretation in (Stapp, 1972)). 
This approach led to the dead-end in the form of slogan:  {\it ``Shut up and calculate!''} (It is commonly  assigned  to Feynman). 
Following Hertz ideas, I claim that the basic  problems of quantum foundations can be resolved only by rejecting the spirit of Copenhagen and creation of a real quantum theory liberated from 
sensational paradigm.   

At  the same time it is important to understand that Einstein and his followers also suffered from ignoring the Bild conception about the meaning of  scientific theory.  
They followed the old-fashioned descriptive understanding of theory and missed to explore the possibilities opened by Hertz 3 statement. Attempts to establish 
one-to-one correspondence between theory and experiment (as, e.g., in Bohmian mechanics) led either to invention of new concepts (such as, e.g., nonlocality in Bohmian mechanics) 
which do not match to `natural concepts' generated by human experience or makes the project too complicated (as in the case of Einstein's attempts to create the classical field model 
matching with micro-phenomena). 

Of course, the main problem is the spirit of Copenhagen. The majority of the quantum community (especially the young generation) 
is oriented to the observational theory - QM. This theory is powerful 
and convenient, but it does not provide the consistent `Bild' of micro-phenomena. The latter is disturbing. Surprisingly, it is disturbing not only for those
who reject the Copenhagen interpretation (or at least understand its restrictive character), but even for its strongest and world's
 famous supporters.\footnote{During the 20 years of V\"axj\"o conferences on quantum foundations, I was 
lucky to meet in the private and relaxing atmosphere many leading experts in quantum theory and experiment, `big names'. 
Surprisingly, practically all  of them dream for a new quantum theory which (soon or later) will replace the present quantum theory. Unfortunately,  people do not like to 
speak openly about their dreams.  (The later is understandable: typically dreams are too private and personal).   Therefore  young researchers live being sure that the present quantum theory 
is the final theory of micro-phenomena.}       

{\bf Measurement problem:} It cannot be solved in the framework of  the observational theory. One has to introduce hidden variables. (Of course, this viewpoint  may be surprising:
one should use unobservable variables to describe generation of outputs of measurement devices.) 
Bell understood the role of hidden variables in description of the process of quantum measurement very well. And he started the right project, but then he was disappointed  
by `nonlocality catastrophe'.\footnote{In spite of the common opinion that Bell `enjoyed' nonlocality, in reality nonlocality came to him as unexpected consequence 
of his  analysis of the EPR-Bohm correlations, see (Bell, 1964, 1987).  }  As was pointed out, the latter is resulted from ignoring the possibility  provided by Hertz 3.  
 
{\bf Acausality of QM.} Von Neumann emphasized acausality of QM (von Neuman, 1955). He also pointed to specialty of quantum randomness, as irreducible randomness. 
(The latter claim is heavily explored in justification of specialty of randomness generated by quantum random generators.)
Acausality of quantum theory is not surprising, generally acausality is a feature of observational theories. One cannot approach causality without transition to the Bild conception.
Thus quantum acausality and specialty of quantum randomness are not the (mystical) physical features of micro-world, but the features of the use of 
Mach's treatment of a physical theory.   

{\bf Perfect correlations.} The EPR correlations (Einstein, Podolsky, \& Rosen, 1935) neither can be explained by the observational theory - without introducing hidden variables. 
 Bell understood this well and his {\it  original Bell inequality} (Bell, 1964) was derived to analyze this problem. However, at that time it was impossible to prepare singlet 
 states with sufficiently high probability and to perform experiments to test the original Bell inequality. Therefore (to establish some relation to experiment) Bell was convinced to proceed with the 
 CHSH-inequality. Later he had never mentioned the original Bell inequality and its the crucial difference from the CHSH-inequality. The latter 
 has nothing to do with the perfect correlations and the EPR-argument (Khrennikov \&  Basieva, 2018). This paper also contains the analysis of the modern experimental situation and 
the novel possibilities to test the original Bell inequality as well as motivation to test it and not the CHSH-inequality.

 {\bf Quantum nonlocality.} It is considered as the most intriguing feature of quantum theory. The nonlocality  prejudice  is so strong, because it is supported by both camps in quantum foundations, 
those who use observational theory (QM) and those who use descriptive theories (such as Bohmian mechanics). In fact, typically two (totally different) 
nonlocalities generated by observational and descriptive theories are identified into aforementioned `quantum nonlocality'. Genuine quantum (observational) nonlocality is encoded  in the 
tensor product structure and the projection postulate. The descriptive nonlocality is encoded in nonlocal equations of motions, such as in Bohmian mechanics, or in violation of Bell type inequalities
(the latter issue is very delicate and  we shall consider it in more detail below). 

{\bf Violation of Bell inequality. } By taking into account the big impact of the debates about the Bell type inequalities, see, for example, 
(Adenier, Fuchs,   \&   Khrennikov, 2007;  Adenier et al., 2008) and its impact to establishing 
the notion of  quantum nonlocality we specially discuss Bell's studies, from the viewpoint of the Bild conception. Bell suffered 
from the same problem as Einstein and Bohm. He took into account Hertz 1,2 statements, but ignored Hertz 3. He also tried to proceed in  the 
 old-fashioned   descriptive framework and {\it to identify the experimental correlations with correlations based on hidden variables}, see (Khrennikov, 2017 a,b, c; Khrennikov and Basieva, 2018) .
De Broglie understood well that such identification has no physical justification  and that the Bell type inequalities cannot be derived for experimental correlations, see (Khrennikov, 2017 a, b).

{\bf Merging QM and general relativity.} In this project the main efforts we set to  `quantize gravity'. It seems that this activity is totally meaningless. One tries to transform 
the descriptive theory into the observational theory. The situation is really paradoxic: one try to collect in one bottle all problems from resulting from 
ignoring Hertz 1, 2 and Herz 3, see (Khrennikov, 2017 d).  It is not surprising 
that it does not work. Merging cannot be approached neither through quantization of gravity nor via naive descriptive `completion' of quantum theory (in the spirit of Einstein or Bohm). 
Both QM and general relative have to be reconsidered from the viewpoint of the Bild conception.   

\subsection{Correspondence between mathematical formalisms  of  theories of different types}
\label{CC}

Since each theory is based on its mathematical formalism, it is useful to establish correspondences between mathematical formalisms of  different types 
of theories representing the same experimental data. The basic elements of the mathematical formalism of a theory $\tau$ are its state space $S_{\tau}$ 
and the space of variables $V_\tau,$ some space (may be very special) of real functions on $S_{\tau}.$ For two theories $\tau_1$ and $\tau_2,$ one can try to establish correspondence
 between their basic elements. This task is not straightforward. 
In particular, the notion of a state is different for different theories, e.g., for  Bild and observational theories $\tau_{B}$ and $\tau_{O}$  (and we shall be interested in establishing 
correspondence between such two types of theories).  A Bild-theory is causal and here the same initial condition 
implies the same consequence. Observational theories are often acausal. And let us consider such a case, i.e., $\tau_{O}$ is acausal.
 It  would be naive to expect that it would be possible to establish straightforward correspondence 
between the state spaces of these theories. Causality is transformed into acausality through consideration of probability distributions. Therefore by establishing 
correspondence between  $\tau_{B}$ and $\tau_{O}$  we have to consider some space (may be very special) of probability distributions $P_B$ 
on  $S_B$ and map it onto the state space of $\tau_{O}.$ (We assume that states of $\tau_{O}$ are interpreted statistically.)   
Then we have to construct two `physically natural maps', 
\begin{equation}
\label{m1}
J:  P_B \to  S_O, \; J^*:  V_B \to  V_O.
\end{equation}
Here `physically natural' means consistent matching with the experimental facts.  Both theories $\tau_{B}$ and $\tau_{O}$ have 
experimental justification through coupling to facts and the correspondence maps have to couple these experimental justifications. 
(We shall illustrate this statement by considering two theories of micro-phenomena,  QM as $\tau_{O}$ and PCSFT 
as $\tau_{B}).$ Of course, theories  $\tau_{B}$ and $\tau_{O}$ can differ by details of experimental justification. Therefore in correspondence
provided by the maps $J, J^*$ some of these details can be ignored.  

Generally these maps are neither one-to-one nor onto. Let us consider this situation in more detail.
\begin{itemize}
\item A cluster of probability distributions on $S_B$  can be mapped into a state from $S_O$
(generally states of  $\tau_{B}$ and probability distributions of such states are unapproachable by $\tau_{O}).$ 
 \item A cluster of variables of $\tau_{B}$ can be mapped into a variable of $\tau_{O}$
 (the observational  description is often operational; it does not distinguish variables of a causal theory).
\item Not all elements of   $S_O$  and  $V_O$  belong to the images  $J (P_B)$ and  $J^*(V_B).$
 (Even observational theory  $\tau_{O}$ can contain its own ideal elements which need not be reflected in $\tau_{B}).$
\end{itemize}

In a Bild theory  $V_B$ is some space of functions on the state space $S_B,$ maps $f: S_B \to \mathbf{R}.$ 
Such theory is causal, the state $\phi$ uniquely determines the values of all physical variables belonging $V_B: \phi \to f(\phi).$

\section{Results: Correspondence between prequantum classical statistical field theory and quantum mechanics}
\label{PQ}

In QM states are given by density operators acting in complex Hilbert space $H$ (endowed with scalar product $\langle \cdot\vert  \cdot \rangle)$ 
and physical variables (observables)  are represented by Hermitian operators in $H.$ Denote the space of density operators by $S_{\rm{QM}}$ and the 
space of Hermitian operators by $V_{\rm{QM}}.$ 

In PCSFT  (Khrennikov,   2007a, b, 2014, 2017c) states are given by vectors of $H$ (in general non-normalized), i.e., 
$S_{\rm{PCSFT}} =H.$   Physical variables  are represented by quadratic forms on $H,$ i.e., maps of the form $f(\phi) =
\langle \phi \vert A \vert  \phi \rangle,$ where $A\equiv A_f$ is a Hermitian operator.
 Denote the space of quadratic forms by the symbol 
$V_{\rm{PCSFT}}.$ Consider the space of probability distributions on $H$ with zero first momentum, i.e., 
\begin{equation}
\label{m3}
\int_H \langle \phi \vert a\rangle d p(\phi)=0
\end{equation}
for any $a\in  H,$ and finite second momentum, i.e., 
\begin{equation}
\label{m4}
{\cal E}_p \equiv \int_H \Vert \phi \Vert^2 d p(\phi) < \infty.
  \end{equation}
  Denote this space of probability distributions by the symbol $P_{\rm{PCSFT}}.$ 
 We remark that, instead of probability distributions, we can consider $H$-valued random vectors with zero mean value and finite second moment:
 $\xi= \xi(\omega),$ where $\omega$ is the chance parameter, such that  $E [\xi]= 0$ and $E[\Vert \xi\Vert^2] < \infty.$ Denote this space by the symbol 
 $R_{\rm{PCSFT}}$
 We remark that if $H$ is finite-dimensional, these are usual complex vector-valued random variables; if   $H$ is 
infinite-dimensional, then the elements of  $R_{\rm{PCSFT}}$ are random fields. For the latter, the basic example is given by the 
choice $H=L_2(\mathbf{R}^n).$ Here each Bild-state $\phi$ is an $L_2$-function, $\phi: \mathbf{R}^n  \mapsto \mathbf{C}.$ 
Hence, each element of  $R_{\rm{PCSFT}}$ can be represented as a function of two variables, $\xi= \xi(x; \omega):$ chance parameter $\omega$ and space coordinates 
$x.$ This is a random field \cite{RF}. We shall use the same terminology, `random fields', even in the finite-dimensional case. 

We remark that, for the state space $H=L_2(\mathbf{R}^n),$ the quantity ${\cal E}_p$ can be represented as
${\cal E}_p = \int_H  {\cal E}(\phi) d p(\phi),$
where 
${\cal E}(\phi) = \Vert \phi\Vert^2 = \int_{\mathbf{R}^n} \vert\phi (x) \vert^2 dx$
is field's energy. Hence, ${\cal E}_p$ is the average of the field energy with respect to the probability distribution $p$ on the space of fields. We can also use the 
random field representation. Let $\xi= \xi(x; \omega)$ be a random field. Then its energy is the random variable 
$
{\cal E}_\xi(\omega)= \int_{\mathbf{R}^n} \vert\xi (x; \omega) \vert^2 dx
$ and ${\cal E}_p$ is the average 
of the latter (here $p$ is the probability distribution of the random field).   

For any $p \in P_{\rm{PCSFT}},$ its (complex) covariance operator $B_p$ is defined by its bilinear (Hermitian) form:
\begin{equation}
\label{m5}
\langle a \vert B_p\vert b\rangle = \int_H \; \langle a\vert \phi  \rangle  \langle \phi \vert b  \rangle \; d p(\phi), \; a, b \in H, 
\end{equation}
or,  for a random vector $\xi,$ we have:   $\langle a \vert B_\xi\vert b\rangle = E  [\langle a\vert \xi \rangle  \langle \xi \vert b  \rangle].$
Generally a probability distribution (a random field) is not determined by its covariance operator (even  under condition of zero average, see (\ref{m3})).
We remark that such complex covariance operator has the same mathematical properties as a density operator, besides normalization by the trace one; it is 
Hermitian, positively semidefinite, and
trace class.
(The latter property is important in the infinite-dimensional case, e.g., for the state space $H=L_2(\mathbf{R}^n)).$

PCSFT (the Bild-type theory)  is connected with QM through the following formula. For $p \in P_{\rm{PCSFT}}$ and $f \in V_{\rm{PCSFT}},$ we have
\begin{equation}
\label{m6}
\langle f \rangle_p = \int_H f(\phi) d p(\phi)= \rm{Tr} B_p A_f ,
\end{equation}
where $A_f=\frac{1}{2} f^{(2)}(0),$ i.e., $f(\phi) = \langle \phi\vert A_f\vert \phi\rangle.$ 
We remark that the covariance operator and the energy average are coupled through the simple formula:   
\begin{equation}
\label{m5e}
 {\cal E}_p= \int_H \Vert \phi \Vert^2 d p(\phi)  = \rm{Tr} B_p;
\end{equation}
in particular, by normalizing the covariance operator of a random field by average of field's energy we obtain a density operator 
 $\rho_p= B_p/{\cal E}_p.$
 
 Let us consider the following maps $J$ and $J^*,$ see (\ref{m1}),  from PCSFT to QM, 
 \begin{equation}
\label{m1q}
J(p)= \rho_p, \; J^*(f)=  A_f .
\end{equation}
This correspondence connects the averages given by the Bild and observational theories:  
\begin{equation}
\label{m6q}
\frac{1}{{\cal E}_p}\langle f \rangle_p  = \rm{Tr} \rho_p A_f ,
\end{equation}
i.e., the QM and PCSFT averages are coupled with the scaling factor which is equal to the inverse of the average energy of the random field.
Thus density operators are normalized  (by average field energy) covariance operators of random fields and the Hermitian operators representing 
quantum observables  correspond to quadratic forms of fields. 
We can also write the relation (\ref{m6q}) in the form: $\langle \frac{f}{{\cal E}_p} \rangle_p  = \rm{Tr} \rho_p A_f.$ 
 If ${\cal E}_p<<1,$ we can consider the quantity $g_p(\phi)\equiv \frac{f(\phi)}{{\cal E}_p}$ as amplification of 
 the PCSFT physical variable $f.$ Thus through coupling with PCSFT we  can be treat QM as an observational theory describing averages of amplified
`subquantum' physical variables.  

In contrast to QM, PCSFT is causal: selection of a vector (`field') $\phi \in H$ determines the values of all PCSFT-variables, quadratic forms  of classical fields:
$\phi \to  \langle \phi\vert A\vert \phi\rangle.$ 

For physical variables, the correspondence  map $J^*$   is one-to-one, but  the map $J$ is not one-to-one. But it is a surjecion, i.e., it is on-to map.  

 \section*{Discussion}
 
 The aim of this paper is to remind to the quantum foundational community studies of Hertz (and Boltzmann)  on the Bild conception of physical theory; 
 especially Hertz analysis of connection between theory and experiment. 
 We emphasize the similarity of the problems discussed by Hertz in the process of transition from Newtonian mechanics to Maxwellian electromagnetism and 
 the problems of interrelation between classical and quantum physical theories (including the problem of hidden variables).  
The Bild conception can be explored to resolve the basic problems of quantum foundations: measurement problem,  acausality  and irreducible quantum randomness,
quantum nonlocality, merging QM and general relativity.
 
As an example of a Bild-type theory preceding QM (the latter is treated as an observational theory), we consider  prequantum classical statistical field theory - PCSFT. 
In contrast to QM, PCSFT is not based solely on the observational data. It contains images which cannot be coupled straightforwardly to data. In particular, the EPR-Bohm 
correlations cannot be identified with the corresponding PCSFT-correlations, although numerically they coincide.  There exists a natural correspondence between the mathematical entities of 
PCSFT and QM, the correspondence is not one-to-one. The same Bild theory can be coupled to a variety of observational theories and a variety of observational theories can represent 
the same experimental data. PCSFT can be coupled not only to QM, but to another observational theory based on threshold detection of random signals, see 
(Khrennikov, 2012). 

Finally, we stress that the Bild conception can be used to develop a consistent theory of quantum(-like)  cognition and interrelation between matter and mind, 
see (Khrennikov, 2010), cf. (Stapp, 2004).   

\section*{Conflict of interest}

I have no  a financial or personal relationship with a third party whose interests could be positively or negatively influenced by the article's content.

{\bf References}

Adenier, G., Fuchs, C.  A.  \&   Khrennikov, A.  (Eds.),  (2007).  {\it Foundations of Probability and Physics-4,}  Conf.  Proc. {\it 889}, Melville, NY: AIP .

Adenier, G., Khrennikov, A. Yu.,  Lahti, P., Manko, V. I.,  \&  Nieuwenhuizen, Th.M. (Eds.), (2008).   {\it  Quantum Theory: 
Reconsideration of Foundations-4},   Conf. Proc. {\it 962}, Melville, NY: AIP. 

Atmanspacher, H.  \&  Primas, H.  (2005).  Epistemic and ontic quantumrealities. In:  G. Adenier   \&  A. Yu. Khrennikov  (Eds.),  {\it Foundations
of Probability and Physics-3}, pp. 49-62,  Conf. Proc. {\it 750}, Melville, NY: AIP.

Bell, J.   (1964).  On the Einstein-Podolsky-Rosen paradox. {\it  Physics,} {\it 1}, 195-200.

Bell, J.  (1987).  {\it  Speakable and unspeakable in quantum mechanics.}  Cambridge: Cambridge Univ. Press.

Bohr, N.   (1935). Can quantum-mechanical description of physical reality be considered complete? {\it  Phys. Rev.}, {\it 48}, 696-702.

Boltzmann, L. (1905). Über die Frage nach der objektiven Existenz der Vorgänge in der unbelebten Natur". In: J.A. Barth (Ed.),   Leipzig: Populäre Schriften.   

Boltzmann, L. (1974) On the development of the methods of theoretical physics in recent times.  In: B. McGuinness  (Ed.),  {\it Theoretical Physics and Philosophical Problems.} 
Vienna Circle Collection, vol 5. Dordrecht: Springer 

De Broglie, L. (1964). {\it  The current interpretation of wave mechanics: a critical study.} Elsevier.

D'Agostino, S. (1992).  Continuity and completeness in physical theory: Schr\"odinger's return to the wave interpretation of quantum mechanics in the
1950's. In:  M. Bitbol \&  O. Darrigol (Eds.),  {\it  E. Schr\"odinger: Philosophy and the Birth of Quantum Mechanics}, pp. 339-360. Gif-sur-Yvette: Editions Frontieres.

Einstein, A.,  Podolsky, B. Rosen, N. (1935). Can quantum-mechanical description of physical reality be considered complete?.
{\it  Phys. Rev.},  47 (10), 777-780.

Einstein, A.  \& Infeld, L. (1961).  {\it  Evolution of physics: The growth of ideas from early concepts to relativity and quanta.} New-York:  Simon  and Schuster.

Hertz, H. (1899). {\it  The principles of mechanics: presented in a new form.} London: Macmillan.

Khrennikov,  A.  (2007a).  Quantum mechanics as the quadratic Taylor approximation of classical mechanics: The finite-dimensional case. {\it Theor. Math. Phys.},   {\it 152},  1111-1121.

Khrennikov, A.  (2007b) To quantum averages through asymptotic expansion of classical averages on infinite-dimensional space. {\it  J. Math. Phys.},  {\it 48}(1), Art. No. 013512.

 Khrennikov, A. (2010). {\it Ubiquitous  quantum structure: from psychology to finances,} Berlin-Heidelberg-New York: Springer.

Khrennikov, A. (2012). Quantum probabilities and violation of CHSH-inequality from classical random signals  and threshold type detection scheme. {\it Prog. Theor. Phys.} {\it  128},   31-58.

Khrennikov,  A. (2014).  {\it  Beyond Quantum}. Singapore:  Pan Stanford Publ..  
 
Khrennikov, A.  (2017a). After Bell.  {\it Fortschritte der Physik (Progress in Physics)}, {\it  65},  1600014.

Khrennikov, A.  (2017b)  Bohr against Bell: complementarity versus nonlocality. {\it Open Phys.}, {\it 15}, 734-738.

Khrennikov, A. (2017c) Quantum epistemology from subquantum ontology: Quantum mechanics from theory of classical random fields. {\it Ann. Phys.},  {\it 377}, 147-163.

Khrennikov,  A.   (2017d). The present situation in quantum theory and its merging with general relativity. {\it Found. Phys.} 47,  1077–1099.

Khrennikov, A.  \& Basieva, I. (2018). Towards experiments to test violation of the original Bell inequality. {\it Entropy}, {\it 20}(4), 280

Miller, A. I. (1984).  {\it Imagery in scientific thought creating 20th-century physics.}  Boston, MA: Birkhäuser. 

Plotnitsky, A. (2006). {\it  Reading Bohr: Physics and philosophy.} Dordrecht: Springer. 

Plotnitsky, A.  (2009).  {\it Epistemology and probability: Bohr, Heisenberg, Schr\"odinger, and the nature of quantum-theoretical thinking.}  Springer: Heidelberg-Berlin-New York. 

 Plotnitsky, A.  (2016).  {\it The principles of quantum theory, from Planck's quanta to the Higgs boson: The nature of quantum reality and the spirit of Copenhagen.}
 Springer: Heidelberg-Berlin-New York. 

Stapp, H. P. (1972). The Copenhagen interpretation. {\it American J. Phys.} {\it 40,} 1098.

Stapp, H. P.  (2004) {\it Mind, matter, and quantum mechanics.} Springer: Heidelberg-Berlin-New York.

 Von Neuman, J.  (1955).  {\it Mathematical foundations of quantum mechanics.}  Princeton: Princeton Univ. Press.

\end{document}